\documentclass[11pt,english]{article}
\usepackage[utf8]{inputenc}
\usepackage{graphicx}
\usepackage{amsmath}
\usepackage{amssymb}
\usepackage{cite}
\usepackage{verbatim}
\usepackage{ragged2e,amsfonts,graphicx,caption,subcaption,xcolor}
\usepackage[left=2cm,right=2cm,top=3cm,bottom=3cm]{geometry}
\usepackage{hyperref}
\hypersetup{colorlinks=false, linkbordercolor={white}}
\makeatletter

\def\etc{\textit{et al. }}

\usepackage{babel}

\begin{document}
\begin{titlepage}
\thispagestyle{empty}

\bigskip

\begin{center}
\noindent{\Large \textbf {Gravitational bending angle with finite distances by Casimir wormholes}}\\

\vspace{0,5cm}

\noindent{Í. D.D. Carvalho${}^{a}$\footnote{e-mail: icarodias@alu.ufc.br}, G. Alencar${}^{ab}$\footnote{e-mail: geova@fisica.ufc.br} and C. R. Muniz${}^{c}$\footnote{e-mail: celio.muniz@uece.br} }

\vspace{0,5cm}

{\it ${}^a$Departamento de Física, Universidade Federal do Ceará-
Caixa Postal 6030, Campus do Pici, 60455-760, Fortaleza, Ceará, Brazil. \\
 }
 {\it ${}^b$International Institute of Physics - Federal University of Rio Grande do Norte, Campus Universitário, Lagoa Nova, Natal, RN 59078-970, Brazil \\
  }
 {\it ${}^c$\!Universidade Estadual do Cear\'a, Faculdade de Educa\c c\~ao, Ci\^encias e Letras de Iguatu, Iguatu-CE, Brazil.}
\end{center}

\vspace{0.3cm}

\begin{abstract}\noindent
    In this paper, we investigate the gravitational bending angle due to the Casimir wormholes, which consider the Casimir energy as the source. Furthermore, some of these Casimir wormholes regard Generalized Uncertainty Principle (GUP) corrections of Casimir energy. We use the Ishihara method for the Jacobi metric, which allows us to study the bending angle of light and massive test particles for finite distances. Beyond the uncorrected Casimir source, we consider many GUP corrections, namely: the Kempf, Mangano and Mann (KMM) model,  the Detournay, Gabriel and Spindel (DGS) model, and the so-called type II model for the GUP principle.  We also find the deflection angle of light and massive particles in the case of the receiver and the source are far away from the lens. In this case, we also compute the optical scalars: convergence and shear for these Casimir wormholes as a gravitational weak lens.
\end{abstract}
\end{titlepage}

\tableofcontents

\section{Introduction}\label{Sec-1}
General Relativity (GR) is the classical field theory that considers gravity as a deformation of spacetime by matter and energy. This curved spacetime determines the particle trajectories \cite{weinberg1972gravitation, landau2, pad, misner1973gravitation}. The GR theory predicted a (correct) deflection angle of light by the Sun twice larger than the Newtonian Gravity. It led to the first of its experimental tests \cite{dyson1920determination}. As proof of its predictive power, we can mention the detection of gravitational waves in 2016 by the VIRGO and LIGO laboratories \cite{Abbott:2016blz} and the first photo of a black hole carried out by the EHT project in 2019 \cite{Akiyama:2019cqa}. The presence of a distribution of matter and energy leads to a modification of the bending of light when compared with the empty spacetime case. The gravitational bending of light by a distribution of matter and energy is called Gravitational Lens (GL) by an analogy with optics (see Refs. \cite{schneider1992gravitational, Wambsganss:1998gg, bartelmann2010gravitational} and references therein).
The study of GL is very important because it is a way to investigate many subjects such as: cosmological parameters \cite{inada2012sloan, suyu2013two, Collett:2014ola}, the discovery of exoplanets \cite{Beaulieu:2006xj, gaudi2008discovery}, the investigation of massive compact halo objects (MACHOs) \cite{Alcock:2000ph}, and the use of strong gravitational lensing to investigate extra dimensions and Kalb-Ramond field \cite{Chakraborty:2016lxo}. Furthermore, a comprehensive analysis of black hole lensing was done in Refs. \cite{Virbhadra:1999nm,Virbhadra:2008ws}.

Wormholes are solutions to Einstein's equations of General Relativity representing connections between two disjoint regions of spacetime \cite{Visser}. The first solutions presented difficulties concerning their traversing \cite{EinsteineRose, Wheeler, Ker1, Ker2, Ker3}. In the '70s, H. G. Ellis and, independently, K. A. Bronikov found an original viable traversable wormhole solution \cite{Ellis, Bronikov}, which was after generalized by M. Morris and K. Thorne \cite{KipThorne}, with E. Teo finding its rotational counterpart \cite{Teo}. In general, there is a need for forms of exotic matter sourcing the wormhole. The understanding of such solutions has became the topic of intense research in recent years\cite{Gao:2016bin}, including the discovery of  humanly transversable wormholes in Ref. \cite{Maldacena:2020sxe} and quantum improvement due to Asymptotically Safety in Quantum Gravity \cite{Moti:2020whf,Alencar:2021enh}.

The Casimir effect, in the vacuum under certain boundary conditions, involves negative energies of quantum fields(\cite{Moste}, and references therein). This theme has raised new insights about the issue if gravity effectively modifies the vacuum energy (and, vice-versa, if this latter gravitates), at least in the weak-field regime. This is presently an object of discussion in the literature \cite{Sorge2,Lima:2019pbo,Lima:2020igm} and of proposes of observational investigations, as in the Archimedes vacuum weight experiment \cite{Calloni}. Concerning wormholes, the study of Casimir plates in the orbit of wormholes can be found in Refs. \cite{wormholecasimir3,wormholecasimir4,Sorge,Alana,Bezerra}. Since Casimir energy is negative (and therefore exotic), it has been speculated as a source to Wormholes a long time ago by Morris and Thorn and some time later by Visser\cite{Morris:1988tu,Visser}. However, a Wormhole shaped solely by the Casimir energy in four dimensions has been found only very recently -- the so-called Casimir wormholes \cite{Garattini:2019ivd}. The result was soon generalized to three dimensions \cite{Alencar:2021ejd} and to D dimensions \cite{Oliveira:2021ypz}. Regarding these objects, the spacetime is curved due to sources of matter described by an energy-momentum tensor whose quantities are the ones that cause the Casimir effect in Minkowski spacetime, but with spherical symmetry.  Beyond this, Casimir energy is a necessary ingredient to obtain the aforementioned humanly transversable wormholes\cite{Maldacena:2020sxe}.  We should point that at very high energy scales, Casimir wormholes can suffer effects of quantum gravity by considering Generalized Uncertainty Principle (GUP) \cite{Jusufi:2020rpw,Sunil}. These theories  correct the Heisenberg principle near the Planck scale by introducing a minimal invariant length in nature (\cite{GUP}, and references therein). In the context of curved spaces, recent phenomenological bounds on parameters of GUP were extracted from gravitational wave events \cite{DAS}. 

In another direction, the Gauss-Bonnet theorem connects differential geometry with topology \cite{do2016differential, klingenberg2013course}. In 2008, Gibbons and Werner used this theorem to propose a method to compute the deflection angle of light for a static and spherically symmetric (SSS) spacetime \cite{gibbons}. In 2016, Ishihara \etc extended this approach to consider finite distances between the system receiver-lens-source \cite{Ishihara:2016vdc}. Finally, Ono \etc generalized the Ishihara method to include stationary backgrounds \cite{ono2017gravitomagnetic, Ono:2018ybw, Ono:2018jrv}. Currently, various papers use the Gauss-Bonnet theorem to investigate the deflection of light in black holes and wormholes \cite{Werner:2012rc, Jusufi:2018jof,Ovgun:2018fnk, Ovgun:2018prw, Ovgun:2019wej}. Furthermore, in Ref. \cite{Gibbons:2015qja}, Gibbons shows that the massive particles free motion in static spacetimes is given by the geodesics of an energy-dependent Riemannian metric on the spatial sections analogous to Jacobi metric in classical dynamics. Crisnejo and Gallo were the first to study the deflection angle of massive particles in a SSS background through Jacobi's metric and the Gauss-Bonnet theorem \cite{Crisnejo:2018uyn}. Then, many works were done using the Crisnejo and Gallo approach. For example, the studying of deflection of massive particles by wormholes \cite{Li:2019vhp}, the deflection of massive particles by Kerr black hole and Teo wormholes \cite{Jusufi:2018kry}, the investigation of rotating naked singularities from Kerr-like wormholes by their deflection angles of massive particles \cite{Jusufi:2018gnz}, the studying of the circular orbit of a particle and the weak lensing effects \cite{Li:2020wvn}, the deflection of massive particles by SSS spacetime in bumblebee gravity \cite{Carvalho:2021jlp} and by a Kerr-like black hole in bumblebee gravity \cite{Li:2020dln}.

In this paper, we propose to apply the Ishihara method to compute the bending of massive particles and light for the Casimir wormholes of Refs. \cite{Garattini:2019ivd,Jusufi:2020rpw}. In principle, these wormholes have (sub)Planckian dimensions, and could only produce measurable shifts in the trajectories of particles as a result of cumulative effects, when they cross cosmological distances and pass nearby myriads of those objects, which would be imbricated in the very fabric of the spacetime. However, we can conceive macroscopic, even astrophysical Casimir wormholes which have evolved, for instance, from inflationary mechanisms. Beyond this, from the theoretical viewpoint, it is interesting to show that these backgrounds provides a finite distance solution for the bending of particles. Thus, we would like to contribute to the literature about the subject with the present study.

The structure of the paper is as follows. In the section \ref{Sec-2}, we present the Casimir wormhole solutions. In the section \ref{Sec-3}, we present the Ishihara method for the deflection of massive particles case. In the section \ref{Sec-4}, we use the Ishihara method to compute the deflection of massive particles for the Casimir wormhole of Garattini\cite{Garattini:2019ivd}, and the travesable wormholes corrected by GUP\cite{Jusufi:2020rpw,Sunil}. Finally, in section \ref{Sec-5}, we summarize our main results.
 
\section{Casimir wormholes with and without GUP corrections - Review}\label{Sec-2}
In this section, we present the line elements found by Refs \cite{Garattini:2019ivd} and \cite{Jusufi:2020rpw}. We will analyze the deflection angle of light considering finite distances for these solutions in the subsequent sections.
Garattini was the first to investigate if the Casimir energy could be the source of a traversable wormhole. He uses the Moris-Thorne wormhole class \cite{Morris:1988cz} in his investigation. So the line element which represents a static and spherically symmetric wormhole is
\begin{equation}\label{EQ01}
ds^2 = - e^{2\Phi(r)} dt^2 + \frac{1}{1-\frac{b(r)}{r}}dr^2+ r^2 (d\theta^2 +\sin^2{\theta}d\phi^2),
\end{equation}
where $\Phi(r)$ and $b(r)$ is the redshift function and the shape function, respectively. Furthermore, the range of the radial coordinate $r$ is $ [r_0, \infty) $, where $r_0$ is the radii of wormhole throat. Using the equation of state $ P = \omega \rho $ and the properties of a traversable wormhole, Garattini finds the line element
\begin{eqnarray}\label{EQ02}
    ds^2 - \left(\frac{r\omega}{\omega r + a}\right)^{\omega-1} dt^2 \frac{dr^2}{1 -\left(1-\frac{1}{\omega}\right)\frac{a}{r}-\frac{a^2}{\omega r^2}} + r^2 (d\theta^2 + \sin^2\theta d\phi^2),
\end{eqnarray}
This line element is associated with Stress-Energy Tensor 
\begin{eqnarray}\label{EQ03}
    T_{\mu\nu} = \frac{r_0^2}{8 \pi Gc^{-4} \omega r^4 }[diag(-1,-\omega,1,1)+(\omega_t(r)-1)diag(0,0,1,1)],
\end{eqnarray}
where $\omega_t(r) = - [\omega^2(4r-r_0)+r_0(4\omega+1)](4(\omega r + r_0))^{-1}$. Note that the Casimir energy obeys the equation of state $P = 3 \rho$, then $\omega=3$ is a special value. Moreover, when $\omega = 1$, the Garattini solution becomes the Ellis-Bronnikov wormhole of Subplanckian size \cite{Ellis:1973yv, Bronnikov:1973fh}. The traversable wormhole found by Garattini has a Planckian size, so this wormhole is traversable but not in practice. For more details about this solution see Ref. \cite{Garattini:2019ivd} and references therein. We also are interested in some wormhole solutions. These solutions are due to GUP corrections of Casimir energy. In Ref. \cite{Jusufi:2020rpw}, Jusufi \textit{et al.} use three GUP models: (1) the Kempf, Mangano, and Mann (KMM) model, (2) the Detournay, Gabriel and Spindel (DGS) model, and (3) the so-called type II model for GUP principle. The four solutions found by Jusifi \textit{et al.} share the same shape function 
\begin{equation}
    b(r) = r_0- \frac{\pi^3}{90}\left(\frac{1}{r}-\frac{1}{r_0}\right) - \frac{\pi^3 D_i \beta}{270}\left(\frac{1}{r^3}-\frac{1}{r_0^3}\right),
\end{equation}
where $\beta$ is the minimum length parameter and $D_i$ can be D1, D2, or D3. These constants $D_i$ are related to each GUP model. The first solution considers a constant redshift function $\Phi(r)= const$, so the line element is 
\begin{equation}
    ds^2 = - dt^2 + \frac{1}{1-\frac{b(r)}{r}}dr^2+ r^2 (d\theta^2 +\sin^2{\theta}d\phi^2),
\end{equation}
where we use $dt^2 = \exp{(2\Phi(r))}dt'^2$ because $\Phi(r)= const$. The second solution considers a redshift function as $\Phi(r) = r_0/r$, so the line element is 
\begin{equation}
    ds^2 = - e^{\frac{2r_0}{r}}dt^2 + \frac{1}{1-\frac{b(r)}{r}}dr^2+ r^2 (d\theta^2 +\sin^2{\theta}d\phi^2),
\end{equation}
The third solution consider a redshift function that obeys $\exp(2\Phi(r))= 1 + \gamma^2/r^2$, so the line element is 
\begin{equation}
    ds^2 = - \left( 1 + \frac{\gamma^2}{r^2}\right)dt^2 + \frac{1}{1-\frac{b(r)}{r}}dr^2+ r^2 (d\theta^2 +\sin^2{\theta}d\phi^2),
\end{equation}
where $\gamma$ is some positive parameter and $r \geq r_0$. The ultimate solution considers a redshift function that obeys $\exp(2\Phi(r))= (1 + \beta D_i/r^2)^{-2(1+1/\omega)}$, the line element is given by 
\begin{equation}
    ds^2 = - \left(1 + \frac{\beta D_i}{r^2}\right)^{-\frac{2}{(1+1/\omega)}} dt^2 + \frac{1}{1-\frac{b(r)}{r}}dr^2+ r^2 (d\theta^2 +\sin^2{\theta}d\phi^2).
\end{equation}
In the next section, we are present the Ishihara method and its application for massive particles when we consider the Jacobi metric. After we present that method, we will compute the deflection angle of massive particles for these five backgrounds. 

\section{Gravitational bending angle  of massive particles in a static and spherically symmetric background - Review}\label{Sec-3}
In this section, we present the Ishihara method \cite{Ishihara:2016vdc} for massive particles through the Jacobi metric \cite{Gibbons:2015qja}. This approach allows us to compute the deflection angle of massive particles considering the finite distances from the source and the receiver to the lens. We consider a static and spherically symmetric spacetime. So, the \textit{ansatz} for SSS line element is
\begin{equation}\label{EQ1}
    ds^2 = -A(r)dt^2 + B(r) dr^2 + r^2(d\theta^2+\sin^2\theta d\phi^2).
\end{equation}
Note that all line elements of the previous section are SSS type. The Jacobi metric associated with the line-element of equation (\ref{EQ1}) is
\begin{equation}\label{EQ2}
    ds^2 = (E^2 - m^2 A(r))\left[\frac{B(r)}{A(r)}dr^2+\frac{r^2}{A(r)}d\phi^2\right],
\end{equation}
where we are considering the equatorial plane $\theta = \pi/2$ \cite{Li:2019vhp}. The axial symmetry presents in the line-element of equation (\ref{EQ2}) provides an angular momentum conserved
\begin{equation}\label{EQ3}
    J \equiv (E^2 - m^2 A(r))\frac{r^2}{A(r)}\frac{d\phi}{ds} = constant.
\end{equation}
So, we can use equations (\ref{EQ2}) and (\ref{EQ3}) to find the trajectory equation of massive particles in SSS spacetime. This way, the trajectory equation is given by
\begin{equation}\label{EQ4}
    (E^2 - m^2 A(r))^2\frac{B(r)}{A(r)} \left(\frac{dr}{ds}\right)^2 = E^2 - \left(m^2 + \frac{J^2}{r^2}\right)A(r).
\end{equation}
We want the deflection angle of massive particles as a function of some parameters of the system constituted by receiver, lens, and source.  These parameters are the velocity of test massive particle $v$, the impact parameter of the trajectory $b$, the properties of background, and the inverse distances of receiver and source to lens $u_R$ and $u_S$, respectively. It's essential to understand the meaning of this $v$. First, this $v$ is the norm of the velocity of a test particle with mass $m$ seen by a static receiver on distance $r_R$ from the lens. Let us clarify the meaning of $v$. We can rewrite equation (\ref{EQ1}) as $ds^2 = -A(r)dt^2 + dl^2$, so 
\begin{eqnarray}
    ds^2 = -A(1-v^2/A)dt^2,    
\end{eqnarray}
where $v^2 \equiv (dl/dt)^2$. Note that the relation of proper time at a point distant $r_R$ from the lens is $ds^2 = -A(r_R)d\tau^2$, and the energy is $E = mA(r)dt/d\tau$. When we consider the identity $$\left(\frac{ds}{dt}\right)^2 = \left(\frac{ds}{d\tau}\right)^2\left(\frac{d\tau}{dt}\right)^2,$$
we can use the definitions above to write the energy as
\begin{eqnarray}\label{energy}
    E = \frac{m A(r)}{\sqrt{1 - v^2/A(r)}}.
\end{eqnarray}
How this energy is a constant of motion we can compute it when $r \to \infty$ (how all Casimir wormhole are asymptotic flat $A(r\to\infty)=1$) and the energy seen by the receiver on $r_R$, then
\begin{eqnarray}
    E = \frac{m A(r_R)}{\sqrt{1 - v^2/A(r_R)}} = \frac{m}{\sqrt{1 - v_\infty^2}}.
\end{eqnarray}
Then the velocity, $v$, of a test particle with mass $m$ seen by a static receiver on distance $r_R$ from the lens is related with $v_\infty$ by 
\begin{eqnarray}\label{velocity}
v^2 = A_R^2 \left[1 - A_R^2(1-v_\infty^2) \right],
\end{eqnarray}
where $A_R \equiv A(r_R)$. We want to consider the movement constants: energy and angular momentum as the function of the asymptotic velocity $v_\infty$ because this is the approach adopted on articles, and the light case can be found when we consider $v_\infty = 1$. To this end, we represent the angular momentum and the energy on asymptotic limit by $J = mv_{\infty}b/\sqrt{1-v_\infty^2}$ and $E = m/\sqrt{1-v_\infty^2}$, respectively \cite{Crisnejo:2018uyn}, note that the velocity seen by a receiver on distance $r_R$ from the lens is given by equation (\ref{velocity}). Also, we make coordinate transformation $u=1/r$. Then, the trajectory equation (\ref{EQ4}) becomes
\begin{equation}\label{EQ5}
    \left(\frac{du}{d\phi}\right)^2 = \frac{1}{A B}\left[\frac{1}{b^2v_\infty^2}-\left(\frac{1-v_\infty^2}{b^2v_\infty^2}+ u^2\right)A\right].
\end{equation}
As well, we can use $E= m/\sqrt{1-v_\infty^2}$ to rewrite the Jacobi metric of equation (\ref{EQ2}). This equation becomes 
\begin{equation}\label{EQ6}
    ds^2 = m^2 \left(\frac{1}{1-v_\infty^2} - A(r)\right)\left[\frac{B(r)}{A(r)}dr^2+\frac{r^2}{A(r)}d\phi^2\right].
\end{equation}
Let us define two unit vectors of the Jacobi metric of equation (\ref{EQ6}). The first one is the unit radial vector $e_{rad}= (\sqrt{A}/(\sqrt{B}\Omega),0)$. The second one is the unit angular vector $e_{ang}=(0, \sqrt{A}/(\Omega r))$, where
\begin{equation}\label{EQ7}
    \Omega \equiv m^2 \left(\frac{1}{1-v_\infty^2} - A(r)\right).
\end{equation}
Furthermore, we define the unit tangent to trajectory vector as
\begin{equation}\label{EQ8}
(K^r,K^\phi) = \frac{J}{\Omega^2}\frac{A}{r^2}\left(\frac{dr}{d\phi},1\right).
\end{equation}

\begin{figure}[h]
    \centering  
    \includegraphics[width=8cm]{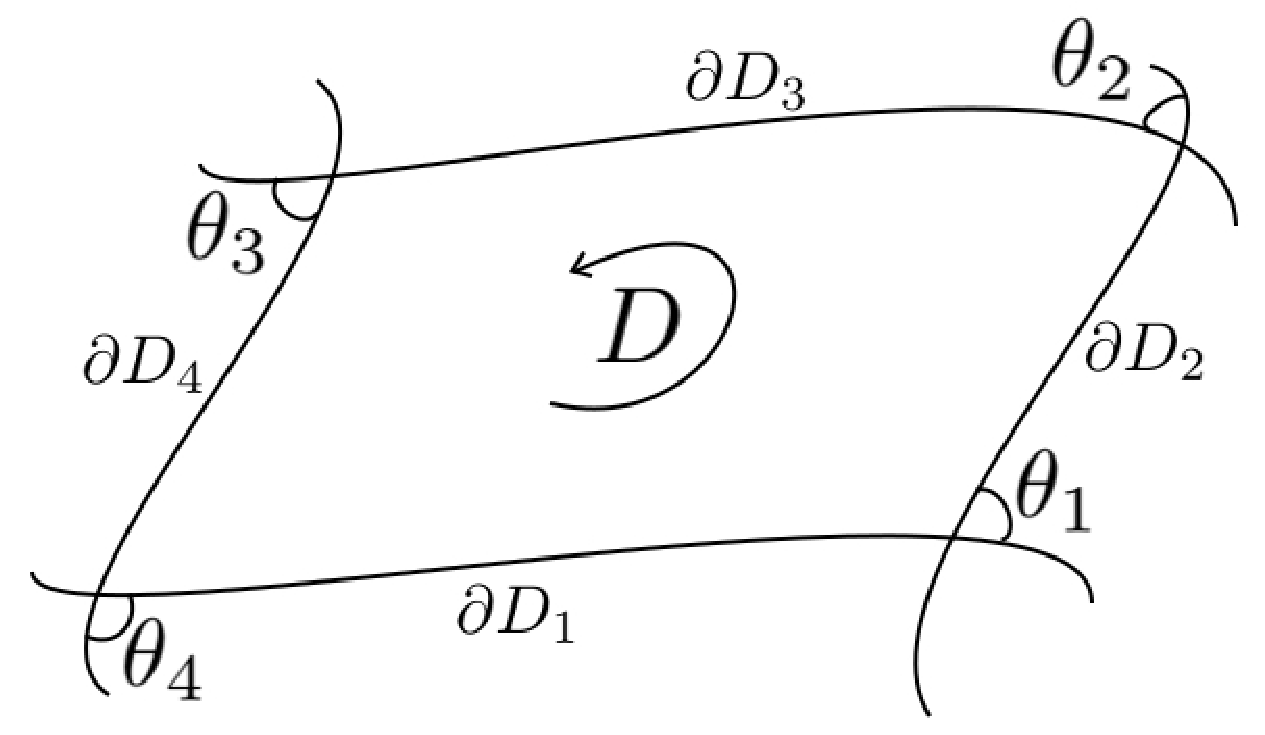}
    \caption{Schematic figure for the Gauss-Bonnet theorem.}
    \label{Fig1}
\end{figure}

The Gauss-Bonnet theorem associates the geometrical characteristics of a surface with its topological Euler characteristic. Let the domain ($D$,$\chi$,$g$) be a subset of a compact, oriented surface, with Euler characteristic $\chi$ and a Riemann metric $g$ giving rise to a Gaussian curvature $K$. Moreover, let $\partial D: \{t\}\rightarrow D$ be a piecewise smooth boundary with geodesic curvature $\kappa_c$, and $\theta_i$ the external angle at $i$th vertex, traversed in the positive sense (see figure \ref{Fig1}). Then the Gauss-Bonnet theorem can be stated as

\begin{equation}\label{EQ9}
    \int\int_{D} K dS + \int_{\partial D} \kappa_c dt + \sum_i \theta_i = 2 \pi \chi(D).
\end{equation}

For more details about the Gauss-Bonnet theorem and its applications see Refs. \cite{do2016differential, klingenberg2013course}. The Ishihara method uses the Gauss-Bonnet theorem to compute the deflection angle of light. We may find the deflection angle of massive particles when we use this method for the Jacobi metric. The Ishihara method writes the gravitational bending angle as
\begin{equation}\label{EQ10}
\alpha = \Psi_R - \Psi_S + \phi_{RS}.
\end{equation}
Let us build the terms that constitute the deflection angle $\alpha$. We can begin with the $\Psi_R - \Psi_S$ term. Let $\Psi(u)$ the angle between the tangent to trajectory vector and the unit radial vector. When we compute the inner product of these vectors, we find
\begin{equation}\label{EQ11}
    \cos{\Psi} = \Omega^2 \left(\frac{B}{A}K^r e_{rad}^r + \frac{r^2}{A}K^\phi e_{rad}^\phi\right) = \sqrt{AB}\frac{J}{\Omega r^2}\frac{dr}{d\phi}.
\end{equation}
If we use the fundamental relation of trigonometry, we avoid the rate $dr/d\phi$, so equation (\ref{EQ11}) provides
\begin{equation}\label{EQ12}
    \sin^2{\Psi} = 1 - \frac{AB}{\Omega^2}J^2\left(\frac{du}{d\phi}\right)^2,
\end{equation}
where we have used $r=1/u$. By using equation (\ref{EQ5}), we obtain
\begin{equation}\label{EQ13}
\Psi(u) = \arcsin{\left(buv_\infty^2 \sqrt{\frac{A}{1- A (1-v_\infty^2)}}~\right)}.
\end{equation}
The $\Psi_R - \Psi_S$ term can be computed with equation (\ref{EQ13}) given $\Psi_R - \Psi_S = \Psi(u_R) + \Psi(u_S) - \pi$. The $\phi_{RS}$ is defined by $\phi_{RS} = \phi(u_R)-\phi(u_S)$, where $\phi(u)$ can be find using equation (\ref{EQ5}). It is important to note that the deflection angle of massive particles becomes the deflection angle of light presented in Ref. \cite{Ishihara:2016vdc} when we consider the velocity of test particle $v_\infty=1$.

Moreover, we are interested in studying some properties of Casimir wormholes as a gravitational lens. Since the deflection angle is not generally observable, we will compute the optical scalars associated with Casimir wormholes. We are studying these scalars on the limit of receptor and source far from the gravitational lens. The weak lens equation relates the source angular position $\beta_i$ (which would be observed if there was not lens) and the actual angular position $\vartheta_i$ is 

\begin{eqnarray}\label{lenseq}
\beta_i = \vartheta_i - \frac{\lambda_{ls}}{\lambda_s} \alpha_i(\vartheta),
\end{eqnarray}
where $\lambda_l$ is the distance from receptor to the lens and $\lambda_{ls}$ is the distance from the lens to the source. The differential of equation (\ref{lenseq}) is $\delta \beta_i = A_{ij} \delta \vartheta_j$, where $A_{ij}$ is the amplification matrix given by
\begin{eqnarray}\label{lensmatrix}
A = \left(
    \begin{matrix}
    1 - \kappa - \tilde{\gamma}_1 & -\tilde{\gamma}_2 - \hat{\omega}\\
    -\tilde{\gamma}_2 + \hat{\omega} & 1 - \kappa +\tilde{\gamma}_1
    \end{matrix}
\right)
\end{eqnarray}
where $\kappa$, $\tilde{\gamma}$ ($\tilde{\gamma}^2 = \tilde{\gamma}^2_1+ \tilde{\gamma}^2_2$), $\hat{\omega}$ are the optical scalars convergence, shear and rotation, respectively \cite{Crisnejo:2017jmx}.
For future motivations, we write the deflection angle of light on the limit of source and receiver far from the lens as $\alpha = \varrho/b + \varsigma/b^2$, when we consider $b = \lambda_l \theta$ ($\lambda_l$ is the distance from receiver to the lens) this deflection angle becomes
\begin{eqnarray}\label{deflectionlens}
    \alpha_i(\theta) = \frac{\varrho}{\lambda_l \theta^2}\theta_i + \frac{\varsigma}{\lambda_l^2 \theta^3}\theta_i,
\end{eqnarray}
The parameters $\varrho$ and $\varsigma$ will be depend of the background. This way, we can find the components of the amplification matrix when we consider the differential of equation (\ref{lenseq}) and equation (\ref{deflectionlens}, then
\begin{eqnarray}\label{matrix}
A_{ij} = \left[1 - \frac{\lambda_{ls}}{\lambda_l}\left(\frac{\varrho}{\lambda_l \vartheta^2}+ \frac{\varsigma}{\lambda_l^2 \vartheta^3}\right)\right]\delta_{ij} +\frac{\lambda_{ls}}{\lambda_l}\left(\frac{2\varrho}{\lambda_l \vartheta^4}+ \frac{3\varsigma}{\lambda_l^2 \vartheta^5}\right)\vartheta_i\vartheta_j.
\end{eqnarray}
Now, we can find the optical scalars through of equations (\ref{lensmatrix}) and (\ref{matrix}). Note that $A_{ij}$ is symmetric, so the optical scalar of rotation is null, $\hat{\omega}=0$. When we consider the trace of the amplification matrix, we find $\mathrm{Tr}(A) = 2(1-\kappa)$, so the convergence is 
\begin{eqnarray}\label{convergence}
    \kappa = - \frac{\lambda_{ls}}{2\lambda_s\lambda^2_{l}} \frac{\varsigma}{\vartheta^3}.
\end{eqnarray}
The shear can be find when we consider the determinant of matrix $\tilde{A}= A - (1-\kappa)\mathbb{I}_2$, where $\mathbb{I}_2$ is the identidy matrix. The shear is $\tilde{\gamma} = \sqrt{\mathrm{det}\tilde{A}}$, we find 
\begin{eqnarray}\label{shear}
    \tilde{\gamma} = \frac{\lambda_{ls}}{\lambda_s}\left(\frac{\varrho}{\lambda_l \vartheta^2}+\frac{3\varsigma}{2\lambda_l^2\vartheta^3}\right).
\end{eqnarray}

In the next section, we will study the deflection angle of massive particles and light and investigate the Casimir wormholes as gravitational lens computing the optical scalars on the limit of receiver and source far from the lens.

\section{Gravitational bending angle of massive particles with correction of finite distances}\label{Sec-4}

In the next subsections, we want to compute the deflection angle of massive particles for five traversable wormhole line-elements. The first solution is a Casimir wormhole, and it was found by Garattini in Ref. \cite{Garattini:2019ivd}. The other solutions are supported by the Casimir energy with GUP corrections\cite{Jusufi:2020rpw}.

\subsection{Casimir wormhole}
Garattini proposed a solution of Casimir wormhole given by line element \cite{Garattini:2019ivd}
\begin{eqnarray}\label{EQ15}
    ds^2 = \left(\frac{r\omega}{\omega r + a}\right)^{\omega-1} dt^2 \frac{dr^2}{1 -\left(1-\frac{1}{\omega}\right)\frac{a}{r}-\frac{a^2}{\omega r^2}} + r^2 (d\theta^2 + \sin^2\theta d\phi^2),
\end{eqnarray}
where the $\omega$ is defined by equation of state $P= \omega \rho$. We are interested in the weak deflection limit, so $a \ll b \leqslant r$, where $b$ is the impact parameter. This way, we can write
\begin{eqnarray}\label{EQ16}
    A(r)^{-1} = \left(\frac{r\omega}{\omega r + a}\right)^{-(\omega-1)} = 1 + \left(1 -\frac{1}{\omega}\right) au + \frac{1}{2}\left(1 - \frac{1}{\omega}\right)\left(1 - \frac{2}{\omega}\right) a^2u^2 + O(a^3),
\end{eqnarray}
where $u = 1/r$. First, we can find $\Psi_R - \Psi_S$ by an expansion of equation (\ref{EQ13}) until first order of $a/b$, where we get

\begin{eqnarray}\label{EQ17}
    \Psi_R - \Psi_S &=& \arcsin(bu_R)+\arcsin(bu_S) - \pi \nonumber \\
    &-& \frac{a}{2v_\infty^2b} \left[\frac{b^2{u_R}^2 c_1}{\sqrt{1-b^2{u_R}^2}}+\frac{b^2{u_S}^2 c_1}{\sqrt{1-b^2{u_S}^2}}\right] \nonumber \\
    &-& \frac{a^2}{b^2}\left[\frac{c_2 b^3{u_R}^3}{2v_\infty^2\sqrt{1-b^2{u_R}^2}}+\frac{c_2 b^3{u_S}^3}{2v_\infty^2\sqrt{1-b^2{u_S}^2}} \right]\nonumber \\
    &+&\frac{a^2}{b^2}\left[\frac{c_1^2b^3{u_R}^3(2b^2{u_R}^2-3)}{8v_\infty^4(1-b^2{u_R}^2)^{3/2}}+\frac{c_1^2b^3{u_S}^3(2b^2{u_S}^2-3)}{8v_\infty^4(1-b^2{u_S}^2)^{3/2}}\right] + O(a^3/b^3),
\end{eqnarray}
where $c_1 \equiv 1-1/\omega$ and $c_2 \equiv (1/2)(1-1/\omega)(1-2/\omega)$. Another term we need calculate, which compound the deflection angle of massive particles, is $\phi_{RS}$. Let's begin finding $u(\phi)$  through the orbit equation (\ref{EQ5}). The orbit equation for the background of equation (\ref{EQ15}) is
\begin{eqnarray}\label{EQ18}
\left(\frac{du}{d\phi}\right)^2 = \left(1 - c_1 au - c_3 a^2u^2\right)\left[\frac{1}{b^2}-u^2+ \frac{c_1 au}{b^2v_\infty^2}+\frac{c_2 a^2u^2}{b^2v_\infty^2}\right],
\end{eqnarray}
where $c_3 \equiv 1/\omega$. We can calulate $u$ iteratively if consider $du/d\phi|_{\phi = \pi/2}=0$. By doing so, we find
\begin{eqnarray}\label{EQ19}
    u = \frac{1}{b}\sin\phi + \frac{ac_1}{2b^2}\left[\frac{1+v_\infty^2}{v_\infty^2}-\sin^2 \phi\right]-a^2\frac{4f_1 + 3f_2}{8b^3}\sin\phi +\frac{a^2f_2}{32b^3}\sin(3\phi),
\end{eqnarray}
where
\begin{eqnarray}\label{EQ20}
    f_1 &\equiv&  \frac{{c_1}^2+2c_2+(3{c_1}^2-2c_3)v_\infty^2}{2v_\infty^2} , \\
    f_2 &\equiv&  \frac{4c_3-3{c_1}^2}{2}.
\end{eqnarray}
Now, we can calculate $ \phi $ iteratively. By doing this, we find
\begin{eqnarray}\label{EQ21}
    \phi = \left\{ \begin{matrix}
        &\phi_0 + a \phi_1 + a^2 \phi_2 + ... ,~~~~~~~~~~~~~~~~~~~~~~&&  if ~|\phi|<\pi/2; \\
        &\pi + \frac{a^2(4f_1+3f_2)\pi}{8b^2}- \phi_0 -a\phi_1-a^2 \phi_2+ ..., && if ~|\phi|>\pi/2,
        \end{matrix}\right.
\end{eqnarray}
where
\begin{eqnarray*}
    \phi_0 &=& \arcsin (bu),\nonumber\\
    \phi_1 &=& -\frac{c_1}{2v_\infty^2b}\frac{(1+v_\infty^2 - b^2u^2v_\infty^2)}{\sqrt{1-b^2u^2}}, \nonumber\\
    \phi_2 &=& \frac{{c_1}^2bu}{8b^2v_\infty^4} \frac{(1+v_\infty^2 - b^2u^2v_\infty^2)^2}{(1-b^2u^2)^{3/2}} - \frac{{c_1}^2 bu}{2v_\infty^2b^2}\frac{(1+v_\infty^2 - b^2u^2v_\infty^2)}{\sqrt{1-b^2u^2}}. \nonumber \\
\end{eqnarray*}
Note that $\phi_R> \pi/2$ and $\phi_S<\pi/2$, so
\begin{eqnarray}\label{EQ22}
    \phi_{RS} &=& \pi -  \arcsin (bu_R) - \arcsin (bu_S)  \nonumber \\
    &+&\frac{a c_1}{2v_\infty^2b}\left[\frac{(1+v_\infty^2 - b^2{u_R}^2v_\infty^2)}{\sqrt{1-b^2{u_R}^2}}+\frac{(1+v_\infty^2 - b^2{u_S}^2v_\infty^2)}{\sqrt{1-b^2{u_S}^2}}\right] \nonumber \\
    &-& \frac{a^2}{b^2}\left[\frac{{c_1}^2bu_R}{8v_\infty^4} \frac{(1+v_\infty^2 - b^2{u_R}^2v_\infty^2)^2}{(1-b^2{u_R}^2)^{3/2}}+\frac{{c_1}^2bu_S}{8v_\infty^4} \frac{(1+v_\infty^2 - b^2{u_S}^2v_\infty^2)^2}{(1-b^2{u_S}^2)^{3/2}}\right] \nonumber \\
    &+&\frac{a^2}{b^2} \left[\frac{{c_1}^2 bu_R}{2v_\infty^2}\frac{(1+v_\infty^2 - b^2{u_R}^2v_\infty^2)}{\sqrt{1-b^2{u_R}^2}}+\frac{{c_1}^2 bu_S}{2v_\infty^2}\frac{(1+v_\infty^2 - b^2{u_S}^2v_\infty^2)}{\sqrt{1-b^2{u_S}^2}} \right]\nonumber \\
    &+& \frac{a^2}{b^2}\left[\frac{4c_3-3{c_1}^2}{64}\right]\left[\frac{bu_R(3-4b^2{u_R}^2)}{\sqrt{1-b^2{u_R}^2}}+\frac{bu_S(3-4b^2{u_S}^2)}{\sqrt{1-b^2{u_S}^2}}\right] \nonumber \\
    &+& \frac{a^2}{b^2}\left[\frac{4{c_1}^2+8c_2+(3{c_1}^2+4c_3)v_\infty^2}{16v_\infty^2}\right]\left[\pi-\arcsin (bu_R) - \arcsin (bu_S)\right]+O(a^3/b^3).
\end{eqnarray}
The deflection angle of massive particles with finite distances by the background of equation (\ref{EQ15}) is
\begin{eqnarray}\label{EQ23}
    \alpha &=& \frac{a^2}{b^2}\left[\frac{4{c_1}^2+8c_2+(3{c_1}^2+4c_3)v_\infty^2}{16v_\infty^2}\right]\left[\pi-\arcsin (bu_R) - \arcsin (bu_S)\right]\nonumber \\
    &+& \frac{a}{b}\frac{ c_1(1+v_\infty^2)}{2v_\infty^2}\left[\sqrt{1-b^2{u_R}^2}+\sqrt{1-b^2{u_S}^2}~\right]\nonumber \\
    &-& \frac{a^2{c_1}^2}{b^2}\left[\frac{bu_R(1+v_\infty^2 - b^2{u_R}^2v_\infty^2)^2-b^3{u_R}^3(2b^2{u_R}^2-3)}{8v_\infty^4(1-b^2{u_R}^2)^{3/2}}+\frac{bu_S(1+v_\infty^2 - b^2{u_S}^2v_\infty^2)^2-b^3{u_R}^3(2b^2{u_R}^2-3)}{8v_\infty^4(1-b^2{u_S}^2)^{3/2}}\right] \nonumber \\
    &+& \frac{a^2}{b^2}\left[\frac{4c_3-3{c_1}^2}{64}\right]\left[\frac{bu_R(3-4b^2{u_R}^2)}{\sqrt{1-b^2{u_R}^2}}+\frac{bu_S(3-4b^2{u_S}^2)}{\sqrt{1-b^2{u_S}^2}}\right] + O(a^3/b^3).
\end{eqnarray}
The deflection angle of light can be obtained by using $v_\infty = 1 $ in equation (\ref{EQ23}). It's important to analyze some regimes of the deflection angle of massive particles. The deflection angle of massive particles in the case of the receiver and source are far way from the lens $(bu_R \approx bu_S \ll 1)$ is
\begin{eqnarray}\label{EQ24}
    {\alpha_\infty} &\approx& \frac{ac_1}{b}\frac{(1+v_\infty^2)}{v_\infty^2} +\frac{a^2\pi}{b^2}\left[\frac{4{c_1}^2+8c_2+(3{c_1}^2+4c_3)v_\infty^2}{16v_\infty^2}\right].
\end{eqnarray}

The deflection of light ($v_\infty=1$) in the limit of equation (\ref{EQ24}) is
\begin{eqnarray}\label{EQ25}
    {\alpha_\infty}_{[light]} &\approx& \frac{2ac_1}{b} +\frac{a^2\pi}{b^2}\left[\frac{4{c_1}^2+8c_2+3{c_1}^2+4c_3}{16}\right].
\end{eqnarray}
When the $P = 3 \rho$ (that is, $\omega = 3$), the constants $c_1=(1-1/\omega)$, $c_2=(1/2)(1-1/\omega)(1-2/\omega)$, $c_3=1/\omega$ become $2/3$, $1/9$, $1/3$, respectively. So, the deflection angle of massive particles with finite distances when $\omega = 3$ is
\begin{eqnarray}\label{EQ231}
    \alpha &=& \frac{a}{b}\frac{(1+v_\infty^2)}{3v_\infty^2}\left[\sqrt{1-b^2{u_R}^2}+\sqrt{1-b^2{u_S}^2}~\right]+ \frac{a^2}{b^2}\left(\frac{1+v_\infty^2}{6v_\infty^2}\right)\left[\pi-\arcsin (bu_R) - \arcsin (bu_S)\right]\nonumber \\
    &-& \frac{4a^2}{9b^2}\left[\frac{bu_R(1+v_\infty^2 - b^2{u_R}^2v_\infty^2)^2-b^3{u_R}^3(2b^2{u_R}^2-3)}{8v_\infty^4(1-b^2{u_R}^2)^{3/2}}+\frac{bu_S(1+v_\infty^2 - b^2{u_S}^2v_\infty^2)^2-b^3{u_R}^3(2b^2{u_R}^2-3)}{8v_\infty^4(1-b^2{u_S}^2)^{3/2}}\right] \nonumber \\
    &+& O(a^3/b^3).
\end{eqnarray}
In the limit case of both source and receiver are very far from the lens, the equation (\ref{EQ231}) becomes
\begin{eqnarray}\label{EQ26}
    {\alpha_\infty} &\approx& \frac{2a}{3b}\frac{(1+v_\infty^2)}{v_\infty^2} +\frac{a^2\pi}{6b^2}\frac{(1+v_\infty^2)}{v_\infty^2}.
\end{eqnarray}
Besides the deflection of massive particles, the deflection of light can be obtained by using of $v_\infty=1$  in equation (\ref{EQ26}), so
\begin{eqnarray}\label{EQ27}
    {\alpha_\infty}_{[light]} &\approx& \frac{4a}{3b}+\frac{a^2\pi}{3b^2}.
\end{eqnarray}
The second term of equation (\ref{EQ27}) disagrees of result obtained by Ref. \cite{Javed:2020mjb}, we believe that this disagreement occurs due to the fact that Ref. \cite{Javed:2020mjb} considers just the trajectory as $u= \frac1b \sin\phi$.
It's interesting to understand the behavior of this Casimir wormhole as a gravitational lens with the deflection angle given by equation (\ref{EQ27}), by comparison of equation (\ref{deflectionlens}), we find $\varrho = 4a/3$ and $\varsigma = \pi a^2/3$. By using the equations (\ref{convergence}) and (\ref{shear}) the optical scalars are
\begin{eqnarray}
    \kappa &=& - \frac{\lambda_{ls}}{\lambda_s\lambda^2_{l}} \frac{\pi a^2}{6}\frac{1}{\vartheta^3},\\
    \tilde{\gamma} &=& \frac{\lambda_{ls}}{\lambda_s}\left(\frac{4a}{3\lambda_l}\frac{1}{\vartheta^2}+\frac{\pi a^2}{2\lambda_l^2}\frac{1}{\vartheta^3}\right), \\
    \hat{\omega} &=& 0.
\end{eqnarray}
These quantities depend of angular throat of wormhole $a$, when $a = 0$ the background becomes Minkowski, the deflection angle is null and the optical scalars are null. In the next subsections, we are going to study the Casimir wormholes with GUP corrections.

\subsection{Model $\Phi = constant$ }\label{Sec-5-1}

The GUP wormhole solution with $\Phi = constant$ is
\begin{equation}\label{EQ28}
ds^2 = - dt^2 + \frac{1}{1 - \frac{r_0}{r}- \frac{\pi^3}{90 r}\left(\frac{1}{r}-\frac{1}{r_0}\right) - \frac{\pi^3 D_i \beta}{270 r}\left(\frac{1}{r^3}-\frac{1}{r_0^3}\right)}dr^2 + r^2 (d\theta^2 +\sin^2{\theta}d\phi^2).
\end{equation}
We can evidence the quantities of interest for the application of the Ishihara method. When we compare equations (\ref{EQ1}) and (\ref{EQ28}), we conclude that $A(r) = 1$ and
\begin{equation}\label{EQ29}
B(r) = \left[1 - \frac{r_0}{r}- \frac{\pi^3}{90 r}\left(\frac{1}{r}-\frac{1}{r_0}\right) - \frac{\pi^3 D_i \beta}{270 r}\left(\frac{1}{r^3}-\frac{1}{r_0^3}\right)\right]^{-1}.
\end{equation}
Rewriting $B(r)$ in power orders of $1/r$, we find
\begin{equation}\label{EQ30}
    B(r) = \left[1 - \left(r_0 - \frac{\pi^3}{90 r_0}-\frac{\pi^3 D_i \beta}{270 r_0^3}\right)\frac{1}{r}-\left(\frac{\pi^3}{90}\right)\frac{1}{r^2}-\left(\frac{\pi^3D_i \beta}{270}\right)\frac{1}{r^4}\right]^{-1}.
\end{equation}
Note that the coefficient of $1/r$ term is the ADM mass present at equation (61) of the Ref. \cite{Jusufi:2020rpw}. We can make some definitions to facilitate our development. Let $\epsilon \equiv \pi^3/90$ and
\begin{equation}\label{EQ31}
\frac{\chi}{b^4} \equiv \frac{\pi^3D_i \beta}{270b^4} = \frac{\epsilon}{90b^2}\frac{D_i \beta}{3b^2},
\end{equation}
where $D_i = \{D_1,D_2,D_3\}$ is a set of numerical factors which depends of the GUP models and $\beta$ is the minimum length parameter \cite{Jusufi:2020rpw}. Note that the $\chi/b^4$ is very lower than $\epsilon/b^2$ and $M/b$ ( $\chi/b^4 \ll \epsilon/b^2$ and $\chi/b^4 \ll M/b$). In this limit, the ADM mass becomes 
\begin{equation}\label{EQ32}
M \equiv r_0 - \frac{\pi^3}{90 r_0}.
\end{equation}
So the equation (\ref{EQ30}) becomes
\begin{equation}\label{EQ33}
B(r) = \left[1 - Mu-\epsilon u^2\right]^{-1},
\end{equation}
Now, let us compute the deflection angle of massive particles present in equation (\ref{EQ10}). We can begin for term $\Psi_R - \Psi_S$. As  $A(r)=1$ then the equation (\ref{EQ13}) becomes $\Psi (u) = \arcsin (bu)$. So, we get
\begin{equation}\label{EQ34}
\Psi_R - \Psi_S = \arcsin (bu_R) + \arcsin (bu_S) - \pi.
\end{equation}\label{EQ35}
To compute the $\phi_{RS}$ term, we are going to use the orbit equation for massive particles, equation (\ref{EQ5}). The orbit equation for the spacetime of equation (\ref{EQ28}) is
\begin{equation}\label{EQ36}
\left(\frac{du}{d\phi}\right)^2 = \left[\frac{1}{b^2}-u^2\right] \left(1 - Mu-\epsilon u^2\right).
\end{equation}
We can calculate $u$ iteratively if consider $du/d\phi |_{\phi = \pi/2} = 0$. By doing so, we find
\begin{equation}\label{EQ37}
u = \frac{1}{b}\sin{\phi} + \frac{M}{2b^2}\cos^2\phi + \frac{\epsilon}{16b^3}\left[-4\phi\cos\phi + \sin(3\phi)\right] + O(M\epsilon,M^2,\epsilon^2).
\end{equation}
Now, we also can calculate $ \phi $ iteratively. This way, $\phi$ is given by
\begin{eqnarray}\label{EQ38}
    \phi = \left\{ \begin{matrix}
        &\phi_0 + M \phi_1 + \epsilon \phi_2+ ... ,~~~~~~~~~~&&  if ~|\phi|<\pi/2; \\
        &\pi + \frac{\pi \epsilon}{4b^2}-\phi_0 - M \phi_1 - \epsilon \phi_2+ ..., && if ~|\phi|>\pi/2,
        \end{matrix}\right.
\end{eqnarray}
where
\begin{eqnarray*}
    \phi_0 &=& \arcsin (bu),\nonumber\\
    \phi_1 &=& -\frac{1}{2b}\sqrt{1-b^2u^2},\nonumber\\
    \phi_2 &=& \frac{1}{4b^2}\arcsin(bu) -\frac{1}{16b^2}\frac{(3-4b^2u^2)bu}{\sqrt{1-b^2u^2}}.\nonumber \\
\end{eqnarray*}
Note that $\phi_R> \pi/2$ and $\phi_S<\pi/2$, so
\begin{eqnarray}\label{EQ39}
    \phi_{RS} &=& \pi + \frac{\pi \epsilon}{4b^2} - [\arcsin(bu_R) +\arcsin(bu_S)]\left(1+ \frac{\epsilon}{4b^2}\right)\nonumber \\
    &+& \frac{M}{b}\left[\sqrt{1-b^2u_R^2}+ \sqrt{1-b^2u_S^2}~\right] \nonumber \\
    &+&\frac{\epsilon}{4b^2}\left[\frac{bu_R(3-4b^2u_R^2)}{\sqrt{1-b^2u_R^2}}+\frac{bu_S(3-4b^2u_S^2)}{\sqrt{1-b^2u_S^2}}\right]+ O(M\epsilon, M^2, \epsilon^2).
\end{eqnarray}
The deflection angle of massive particles for this background can be found by using equations (\ref{EQ34}) and (\ref{EQ39}) at equation (\ref{EQ10}). The deflection angle considering finite distances is
\begin{eqnarray}\label{EQ40}
\alpha &=& \frac{\pi \epsilon}{4b^2} - [\arcsin(bu_R) +\arcsin(bu_S)]\frac{\epsilon}{4b^2}\nonumber \\
&+& \frac{M}{2b}\left[\sqrt{1-b^2u_R^2}+ \sqrt{1-b^2u_S^2}~\right] \nonumber \\
&+&\frac{\epsilon}{4b^2}\left[\frac{bu_R(3-4b^2u_R^2)}{\sqrt{1-b^2u_R^2}}+\frac{bu_S(3-4b^2u_S^2)}{\sqrt{1-b^2u_S^2}}\right]+ O(M\epsilon, M^2, \epsilon^2).
\end{eqnarray}
First, we can see that velocity $v$ does not appear in equation (\ref{EQ40}). This is a consequence of $A(r)=1$. In this case, the deflection angle of massive particles equals the deflection angle of light. Second, we can find the deflection angle of light in the case of the receiver and source are far way of the lens ($bu_R \approx bu_S \ll 1$), the deflection of light in this case is
\begin{eqnarray}\label{EQ41}
\alpha_\infty &\approx& \frac{r_0}{b} - \frac{\pi^3}{90 r_0b}\left(1 - \frac{\pi r_0}{4b}\right),
\end{eqnarray}
where we use the definition $\epsilon \equiv \pi^3/90$ and the definition of $M$ present in equation (\ref{EQ32}).
The equation (\ref{EQ41}) agrees with Ref. \cite{Jusufi:2020rpw}. In the next subsections, we consider the  three solutions with $A(r) \neq 1$.
How we done before, we are going to compute the optical scalars related to this wormhole as a gravitational lense. By comparison of equations (\ref{deflectionlens}) and (\ref{EQ41}), we find $\varrho = r_0 - \pi/(90r_0)$ and $\varsigma = \pi^4/360$. By using the equations (\ref{convergence}) and (\ref{shear}) the optical scalars are
\begin{eqnarray}
    \kappa &=& - \frac{\lambda_{ls}}{\lambda_s\lambda^2_{l}} \frac{\pi^4}{720}\frac{1}{\vartheta^3},\\
    \tilde{\gamma} &=& \frac{\lambda_{ls}}{\lambda_s}\left(\frac{1}{\lambda_l}\left(r_0 - \frac{\pi}{90 r_0}\right)\frac{1}{\vartheta^2}+\frac{1}{\lambda_l^2}\frac{\pi^4}{240}\frac{1}{\vartheta^3}\right), \\
    \hat{\omega} &=& 0.
\end{eqnarray}
In the next subsection, we are going to study de deflection angle of massive particles and light by the Casimir wormholes with variavel redshift function $\Phi(r)$.
\subsection{Model $ \Phi = \frac{r_0}{r}$}\label{Sec-5-2}

The GUP wormhole solution with $\Phi = \frac{r_0}{r}$ is
\begin{equation}\label{EQ42}
ds^2 = - e^{\frac{2r_0}{r}}dt^2 + \frac{1}{1 - \frac{r_0}{r}- \frac{\pi^3}{90 r}\left(\frac{1}{r}-\frac{1}{r_0}\right) - \frac{\pi^3 D_i \beta}{270 r}\left(\frac{1}{r^3}-\frac{1}{r_0^3}\right)}dr^2+r^2 (d\theta^2 +\sin^2{\theta}d\phi^2).
\end{equation}

In this subsection, we use the same approach as the subsection above to find the deflection angle of massive particles. We can begin by computing the $\Psi_R - \Psi_S$ term of equation (\ref{EQ10}). We can do this by finding $\Psi(u)$ through equation (\ref{EQ13}). The equation (\ref{EQ13}) provides
\begin{equation}\label{EQ43}
    \Psi(u) = \arcsin\left(buv_\infty \left(e^{-2r_0/r}-(1-v_\infty^2)\right)^{-1/2}\right).
\end{equation}
We consider the weak deflection limit $r_0 \ll b \leqslant r$ then we can write $e^{-2r_0/r}\approx 1 - 2r_0/r$, so making an expansion we get
\begin{equation}\label{EQ44}
    \Psi(u) = \arcsin(bu) + \frac{r_0}{bv_\infty^2}\frac{b^2u^2}{\sqrt{1-b^2u^2}} + O(r_0^2/b^2).
    \end{equation}
This way, the $\Psi_R-\Psi_S$ term of equation (\ref{EQ10}) is
\begin{equation}\label{EQ45}
    \Psi_R-\Psi_S = -\pi + \arcsin(bu_R)+\arcsin(bu_S) + \frac{r_0}{bv_\infty^2}\left[\frac{b^2{u_R}^2}{\sqrt{1-b^2{u_R}^2}}+\frac{b^2{u_S}^2}{\sqrt{1-b^2{u_S}^2}}\right] + O(r_0^2/b^2).
\end{equation}
As was done the previous subsection,  we are going to compute the $\phi_{RS}$ term by using of the orbit equation for massive particles, equation (\ref{EQ5}). The orbit equation for the line element of equation (\ref{EQ42}) is
\begin{equation}\label{EQ46}
    \left(\frac{du}{d\phi}\right)^2 = \left[\frac{1}{b^2}-u^2 - \frac{2r_0u}{b^2v_\infty^2}\right] \left(1 - Mu-\epsilon u^2\right),
\end{equation}
where $\epsilon \equiv \pi^3/90$ and the definition of $M$ present in equation (\ref{EQ32}). As well as in the above subsection, we can calculate $u$ iteratively if consider $du/d\phi |_{\phi = \pi/2} = 0$. By doing so, we find
\begin{equation}\label{EQ47}
    u = \frac{1}{b}\sin{\phi} + \frac{M}{2b^2}\cos^2\phi -\frac{r_0}{b^2v_\infty^2} + \frac{\epsilon}{16b^3}\left[-4\phi\cos\phi + \sin(3\phi)\right]+ O\left(Mr_0,M\epsilon,r_0\epsilon,M^2,r_0^2,\epsilon^2\right).
\end{equation}
Now, we also can calculate $\phi$ iteratively. This way, $\phi$ is given by
\begin{eqnarray}\label{EQ48}
    \phi = \left\{ \begin{matrix}
        &\phi_0 + M \phi_{1,1} + r_0 \phi_{1,2}+ \epsilon \phi_2+ ... ,~~~~~~~~~~&&  if ~|\phi|<\pi/2; \\
        &\pi + \frac{\epsilon \pi}{4b^2}-\phi_0 - M \phi_{1,1} - r_0 \phi_{1,2}- \epsilon \phi_2+ ..., && if ~|\phi|>\pi/2,
        \end{matrix}\right.
\end{eqnarray}
where
\begin{eqnarray*}
    \phi_0 &=& \arcsin (bu),\nonumber\\
    \phi_{1,1} &=& -\frac{1}{2b}\sqrt{1-b^2u^2},\nonumber\\
    \phi_{1,2} &=& \frac{1}{bv_\infty^2} \frac{1}{\sqrt{1-b^2u^2}}, \nonumber\\
    \phi_2 &=& \frac{1}{16b^2}\left(4\arcsin(bu)-\frac{(3-4b^2u^2)bu}{\sqrt{1-b^2u^2}}\right).\nonumber \\
\end{eqnarray*}
As $\phi_R> \pi/2$ and $\phi_S<\pi/2$, we can find $\phi_{RS}$. We obtain
\begin{eqnarray}\label{EQ49}
\phi_{RS} &=& \pi + \frac{\epsilon \pi}{4b^2} -\left(1+\frac{\epsilon}{4b^2}\right)\left[\arcsin(bu_R)+\arcsin(bu_S)\right] \nonumber \\
&+& \frac{M}{2b} \left[\sqrt{1-b^2u_R^2}+\sqrt{1-b^2u_S^2}\right] - \frac{r_0}{bv_\infty^2}\left[\frac{1}{\sqrt{1-b^2u_R^2}}+\frac{1}{\sqrt{1-b^2u_S^2}}\right] \nonumber \\
&+&\frac{\epsilon}{16b^2}\left[\frac{(3-4b^2u_R^2)bu_R}{\sqrt{1-b^2u_R^2}}+\frac{(3-4b^2u_S^2)bu_S}{\sqrt{1-b^2{u_S}^2}}\right]+ O\left(Mr_0,M\epsilon,r_0\epsilon,M^2,r_0^2\right).
\end{eqnarray}
The deflection angle of massive particles for this background can be found by using equations (\ref{EQ45}) and (\ref{EQ49}) at equation (\ref{EQ10}). The deflection angle considering finite distances is
\begin{eqnarray}\label{EQ50}
\alpha &=&\frac{\epsilon \pi}{4b^2} -\frac{\epsilon}{4b^2}\left[\arcsin(bu_R)+\arcsin(bu_S)\right]+ \left[\frac{M}{2b}- \frac{r_0}{bv_\infty^2}\right] \left[\sqrt{1-b^2u_R^2}+\sqrt{1-b^2u_S^2}\right] \nonumber \\
&+&\frac{\epsilon}{16b^2}\left[\frac{(3-4b^2u_R^2)bu_R}{\sqrt{1-b^2u_R^2}}+\frac{(3-4b^2u_S^2)bu_S}{\sqrt{1-b^2{u_S}^2}}\right]+ O\left(Mr_0,M\epsilon,r_0\epsilon,M^2,r_0^2\right).
\end{eqnarray}
This result agrees as equation (\ref{EQ40}) when $A(r)=1$ (when we do not consider the term $r_0/bv_\infty^2$ in equation (\ref{EQ50})). The deflection angle of light can be obtained when we consider $v_\infty = 1 $ on equation  (\ref{EQ50}). In the case of the receiver and source are far way from the lens $(bu_R \approx bu_S \ll 1)$, the deflection angle of massive particles is
\begin{eqnarray}\label{EQ51}
\alpha_\infty \approx -\frac{r_0}{b}\left(\frac{2-v_\infty^2}{v_\infty^2}\right) - \frac{\pi^3}{90br_0}\left(1 - \frac{\pi r_0}{4b}\right),
\end{eqnarray}
where we use the definition $\epsilon \equiv \pi^3/90$ and the definition of $M$ present in equation (\ref{EQ32}). Finally, for compute the deflection angle of light in this limit we just use $v_\infty=1$ in equation (\ref{EQ51}). Then
\begin{eqnarray}\label{EQ52}
    {\alpha_\infty}_{[light]} \approx -\frac{r_0}{b} - \frac{\pi^3}{90br_0}\left(1 - \frac{\pi r_0}{4b}\right).
\end{eqnarray}
This result agrees with Ref. \cite{Jusufi:2020rpw} only for term $-r_0/b$. The negative value of the results of equations (\ref{EQ51}) and (\ref{EQ52}) refer to the moving away of the particles when deflected by the wormhole of equation (\ref{EQ42}). Of course, the resulting negative value should be taken as an absolute value $|\alpha|$. Taking the absolute value of the deflection angle of equation (\ref{EQ52}) and comparing it with equation (\ref{deflectionlens}), we find $\varrho = r_0 + \pi/(90r_0)$ and $\varsigma = -\pi^4/360$. By using the equations (\ref{convergence}) and (\ref{shear}) the optical scalars are
\begin{eqnarray}
    \kappa &=& \frac{\lambda_{ls}}{\lambda_s\lambda^2_{l}}\frac{\pi^4}{720} \frac{1}{\vartheta^3},\\
    \tilde{\gamma} &=& \frac{\lambda_{ls}}{\lambda_s}\left(\frac{1}{\lambda_l}\left(r_0 + \frac{\pi}{90 r_0}\right)\frac{1}{\vartheta^2}-\frac{1}{\lambda_l^2}\frac{\pi^4}{240}\frac{1}{\vartheta^3}\right), \\
    \hat{\omega} &=& 0.
\end{eqnarray}
However, the optical scalars for this lens are similar to the ones of the $\Phi(r)= const$ they do not equal, this way the images provide by these backgrounds are different.

\subsection{Model $e^{2\Phi(r)} = 1 + \frac{\gamma^2}{r^2}$}\label{Sec-5-3}

The GUP wormhole solution with $e^{2\Phi(r)} = 1 + \frac{\gamma^2}{r^2}$ is
\begin{equation}\label{EQ53}
ds^2 = - \left(1 + \frac{\gamma^2}{r^2}\right) dt^2 + \frac{1}{1 - \frac{r_0}{r}- \frac{\pi^3}{90 r}\left(\frac{1}{r}-\frac{1}{r_0}\right) - \frac{\pi^3 D_i \beta}{270 r}\left(\frac{1}{r^3}-\frac{1}{r_0^3}\right)}dr^2+r^2 (d\theta^2 +\sin^2{\theta}d\phi^2).
\end{equation}
Now, let us compute the deflection angle of massive particles and light for the solution present in equation (\ref{EQ53}). As was done in the subsections above, we can begin by computing the deflection angle of massive particles by term $\Psi_R-\Psi_S$ of equation (\ref{EQ10}). The equation (\ref{EQ13}) provides
\begin{equation}\label{EQ54}
\Psi(u) = \arcsin\left(buv \left(\left(1+\gamma^2u^2\right)^{-1}-(1-v_\infty^2)\right)^{-1/2}\right).
\end{equation}
We are interested in the weak deflection limit. Thus it is reasonable to consider $\gamma^2/r^2<\gamma^2/b^2 \ll 1$, so making an expansion we get
\begin{equation}\label{EQ55}
\Psi(u) = \arcsin\left(bu\right) + \frac{\gamma^2}{2v_\infty^2b^2}\frac{b^3u^3}{\sqrt{1-b^2u^2}}+O\left(\frac{\gamma^4}{b^4}\right).
\end{equation}
Then, the term $\Psi_R - \Psi_S$ becomes
\begin{equation}\label{EQ56}
    \Psi_R-\Psi_S = \arcsin\left(bu_R\right)+\arcsin\left(bu_S\right)-\pi + \frac{\gamma^2}{2v_\infty^2b^2}\left[\frac{b^3u_R^3}{\sqrt{1-b^2u_R^2}}+\frac{b^3u_S^3}{\sqrt{1-b^2u_S^2}}\right]+O\left(\frac{\gamma^4}{b^4}\right).
\end{equation}
Now, we are going to compute the $\phi_{RS}$ term of equation (\ref{EQ10}). We can do this through the orbit equation, which for the solution present in equation  (\ref{EQ53}) is
\begin{equation}\label{EQ57}
\left(\frac{du}{d\phi}\right)^2 = \left[\frac{1}{b^2}-u^2-\frac{\gamma^2u^2}{b^2v_\infty^2}\right]\left(1-Mu-\epsilon u^2\right),
\end{equation}
where we consider $(1+\gamma^2u^2)^{-1}\approx 1-\gamma^2u^2$ because we are interested in the first-order of the bending angle. So, computing $u$ iteratively  when consider $du/d\phi|_{\phi=\pi/2}=0$ and computing $\phi$ with same approach, we got
\begin{eqnarray}\label{EQ58}
    \phi = \left\{ \begin{matrix}
        &\phi_0 + M \phi_1+ \gamma^2\phi_{2,1} + \epsilon \phi_{2,2}+ ... ,~~~~~~~~~~&&  if ~|\phi|<\pi/2; \\
        &\pi + \frac{\epsilon \pi}{4b^2} +\frac{\pi\gamma^2}{2b^2v_\infty^2}-\phi_0 - M \phi_1- \gamma^2\phi_{2,1} - \epsilon \phi_{2,2}+ ..., && if ~|\phi|>\pi/2,
    \end{matrix}\right.
\end{eqnarray}
where
\begin{eqnarray*}
    \phi_0 &=& \arcsin (bu),\nonumber\\
    \phi_{1} &=& -\frac{1}{2b}\sqrt{1-b^2u^2},\nonumber\\
    \phi_{2,1} &=&  \frac{1}{2b^2v_\infty^2}\arcsin(bu), \nonumber\\
    \phi_{2,2} &=& \frac{1}{16b^2}\left(4\arcsin(bu)-\frac{(3-4b^2u^2)bu}{\sqrt{1-b^2u^2}}\right).\nonumber \\
\end{eqnarray*}
As $\phi_S < \pi/2$ and $\phi_R> \pi/2$, we can write
\begin{eqnarray}\label{EQ59}
\phi_{RS} &=& \pi + \frac{\epsilon \pi}{4b^2}+\frac{\pi\gamma^2}{2b^2v_\infty^2} -\left(1+\frac{\epsilon}{4b^2}+\frac{\gamma^2}{2b^2v_\infty^2}\right)\left[\arcsin(bu_R)+\arcsin(bu_S)\right] \nonumber \\
&+& \frac{M}{2b} \left[\sqrt{1-b^2u_R^2}+\sqrt{1-b^2u_S^2}\right]+\frac{\epsilon}{16b^2}\left[\frac{(3-4b^2u_R^2)bu_R}{\sqrt{1-b^2u_R^2}}+\frac{(3-4b^2u_S^2)bu_S}{\sqrt{1-b^2{u_S}^2}}\right] \nonumber \\
&+& O\left(M\gamma^2,M\epsilon,\gamma^2\epsilon,M^2,\gamma^4\right).
\end{eqnarray}
We can get the deflection angle of massive particles by using the equations (\ref{EQ56}) and (\ref{EQ59}) in (\ref{EQ10}). The deflection angle of massive particles is
\begin{eqnarray}\label{EQ60}
\alpha &=&\frac{\epsilon \pi}{4b^2} +\frac{\pi\gamma^2}{2b^2v_\infty^2}-\left(\frac{\epsilon}{4b^2}+\frac{\gamma^2}{2b^2v_\infty^2}\right)\left[\arcsin(bu_R)+\arcsin(bu_S)\right] \nonumber \\
&+& \frac{M}{2b}\left[\sqrt{1-b^2u_R^2}+\sqrt{1-b^2u_S^2}\right] +\frac{\gamma^2}{2b^2v_\infty^2}\left[\frac{b^3u_R^3}{\sqrt{1-b^2u_R^2}}+\frac{b^3u_S^3}{\sqrt{1-b^2u_S^2}}\right] \nonumber \\
&+&\frac{\epsilon}{16b^2}\left[\frac{(3-4b^2u_R^2)bu_R}{\sqrt{1-b^2u_R^2}}+\frac{(3-4b^2u_S^2)bu_S}{\sqrt{1-b^2{u_S}^2}}\right]+  O\left(M\gamma^2,M\epsilon,\gamma^2\epsilon,M^2,\gamma^4\right).
\end{eqnarray}
This result agrees to equation (\ref{EQ40}) when $A(r)=1$ (when we consider $\gamma^2 = 0$ in equation (\ref{EQ53})). The deflection angle of light can get by considering $v_\infty=1$. In the case of the receiver and source are far way from the lens $(bu_R \approx bu_S \ll 1)$, the deflection angle of massive particles is
\begin{eqnarray}\label{EQ61}
{\alpha_\infty} &\approx& \frac{r_0}{b} - \frac{\pi^3}{90 r_0b}\left(1 - \frac{\pi r_0}{4b}\right) +\frac{\pi\gamma^2}{2b^2v_\infty^2},
\end{eqnarray}
where we used the definition $\epsilon \equiv \pi^3/90$ and the definition of $M$ present in equation (\ref{EQ32}). The deflection angle of light, in this case, can be found by using $v_\infty=1$ in equation (\ref{EQ61}). The equation (\ref{EQ61}) agrees with Ref. \cite{Jusufi:2020rpw} only at two first terms when we consider $v_\infty=1$. The $\gamma^2$ term disagrees with the potency of $\gamma$ (in Ref. \cite{Jusufi:2020rpw} this term is dimensionally wrong) and with the sign.

Now, we can find the optical scalars for this background when it is a gravitational lens that has a deflection angle of light given by equation (\ref{EQ61}) whith $v_\infty = 1$. When we compare the equations (\ref{EQ61}) with $v_\infty=1$ and (\ref{deflectionlens}), we find $\varrho = r_0 - \pi/(90r_0)$ and $\varsigma = \pi^4/360+\pi\gamma^2/2$. By using the equations (\ref{convergence}) and (\ref{shear}) the optical scalars are
\begin{eqnarray}\label{conv_gamma}
    \kappa &=& - \frac{\lambda_{ls}}{2\lambda_s\lambda^2_{l}} \left(\frac{\pi^4}{360}+\frac{\pi \gamma^2}{2}\right)\frac{1}{\vartheta^3},\\\label{shear_gamma}
    \tilde{\gamma} &=& \frac{\lambda_{ls}}{\lambda_s}\left(\frac{1}{\lambda_l}\left(r_0 - \frac{\pi}{90 r_0}\right)\frac{1}{\vartheta^2}+\frac{3}{2\lambda_l^2}\left(\frac{\pi^4}{360}+\frac{\pi \gamma^2}{2}\right)\frac{1}{\vartheta^3}\right), \\
    \hat{\omega} &=& 0.
\end{eqnarray}
Note that these optical scalars are different from the others, so this background generates a different image of the same source.
\subsection{ Isotropic model with $\omega_r (r)= constant$}\label{Sec-5-4}

Isotropic model with is $e^{2\Phi(r)} = \left(1 + \frac{\beta D_i}{r^2}\right)^{- \frac{2}{1+1/\omega}}$
\begin{equation}\label{EQ-5-4-1}
ds^2 = - \left(1 + \frac{\beta D_i}{r^2}\right)^{- \frac{2}{1+1/\omega}} dt^2 + \frac{1}{1 - \frac{r_0}{r}- \frac{\pi^3}{90 r}\left(\frac{1}{r}-\frac{1}{r_0}\right) - \frac{\pi^3 D_i \beta}{270 r}\left(\frac{1}{r^3}-\frac{1}{r_0^3}\right)}dr^2 + r^2 (d\theta^2 +\sin^2{\theta}d\phi^2)
\end{equation}
The computation of deflection angle of massive particles is very similar to the solution of the subsection above, where
\begin{equation}\label{EQ-5-4-2}
    A^{-1} \approx 1-\gamma^2 u^2
\end{equation}
because for the solution of equation (\ref{EQ-5-4-1}), in the weak deflection limit, we have
\begin{equation}\label{EQ-5-4-3}
    A^{-1} =\left(1 + \beta D_iu^2\right)^{\frac{2}{1+1/\omega}} \approx 1-\left(-\frac{2}{1+\omega^{-1}}\beta D_i \right) u^2.
\end{equation}
So, in first-order, the deflection angle of massive particles is the same as equation (\ref{EQ60}) when
\begin{equation}\label{EQ-5-4-4}
\gamma^2 = -\frac{2}{1+\omega^{-1}}\beta D_i.
\end{equation}
Thus,  the deflection of massive particles is
\begin{eqnarray}\label{EQ-5-4-5}
\alpha &=&\frac{\epsilon \pi}{4b^2} - \frac{1}{1+\omega^{-1}} \frac{\pi\beta D_i}{b^2v_\infty^2}-\left(\frac{\epsilon}{4b^2}-\frac{1}{1+\omega^{-1}} \frac{\beta D_i}{b^2v_\infty^2}\right)\left[\arcsin(bu_R)+\arcsin(bu_S)\right] \nonumber \\
&+& \frac{M}{2b}\left[\sqrt{1-b^2u_R^2}+\sqrt{1-b^2u_S^2}\right] -\frac{1}{1+\omega^{-1}} \frac{\beta D_i}{b^2v_\infty^2}\left[\frac{b^3u_R^3}{\sqrt{1-b^2u_R^2}}+\frac{b^3u_S^3}{\sqrt{1-b^2u_S^2}}\right] \nonumber \\
&+&\frac{\epsilon}{16b^2}\left[\frac{(3-4b^2u_R^2)bu_R}{\sqrt{1-b^2u_R^2}}+\frac{(3-4b^2u_S^2)bu_S}{\sqrt{1-b^2{u_S}^2}}\right]+  O\left(M\gamma^2,M\epsilon,\gamma^2\epsilon,M^2,\gamma^4\right).
\end{eqnarray}
The deflection angle of light can get by considering $v_\infty=1$. The deflection angle of massive particles in the case of the receiver and source were a far way of the lens $(bu_R \approx bu_S \ll 1)$ is
\begin{eqnarray}\label{EQ-5-4-6}
\alpha_\infty &\approx& \frac{r_0}{b} - \frac{\pi^3}{90 r_0b}\left(1 - \frac{\pi r_0}{4b}\right) -\frac{1}{1+w^{-1}}\frac{\pi\beta D_i}{b^2v_\infty^2}.
\end{eqnarray}
The deflection angle of light ($v_\infty=1$) when we consider the receiver and source very far away of the lens and $\omega = 1$ is
\begin{eqnarray}\label{EQ-5-4-7}
{\alpha_\infty}_{[light]} &\approx& \frac{r_0}{b} - \frac{\pi^3}{90 r_0b}\left(1 - \frac{\pi r_0}{4b}\right) -\frac{\pi\beta D_i}{2b^2}.
\end{eqnarray}
This result agrees with Ref. \cite{Jusufi:2020rpw} only at the two firsts terms and disagrees with the last term due to the sign.
We also can compute the optical scalars for this background, we can use the equation (\ref{EQ-5-4-4}) in (\ref{conv_gamma}) and (\ref{shear_gamma}), 
\begin{eqnarray}
    \kappa &=& - \frac{\lambda_{ls}}{2\lambda_s\lambda^2_{l}} \left(\frac{\pi^4}{360}-\frac{\pi\omega}{1+\omega}\beta D_i\right)\frac{1}{\vartheta^3},\\
    \tilde{\gamma} &=& \frac{\lambda_{ls}}{\lambda_s}\left(\frac{1}{\lambda_l}\left(r_0 - \frac{\pi}{90 r_0}\right)\frac{1}{\vartheta^2}+\frac{3}{2\lambda_l^2}\left(\frac{\pi^4}{360}-\frac{\pi\omega}{1+\omega}\beta D_i\right)\frac{1}{\vartheta^3}\right), \\
    \hat{\omega} &=& 0.
\end{eqnarray}
These are the optical scalars of this background. The modification of the source image is given by these quantities.

\section{Conclusion}\label{Sec-5}
In this paper, we computed the deflection angle of massive particles and light by some Casimir wormholes. We used Ishihara's method because it provides a deflection angle that depends on distances from the source to the lens and from this latter to the receiver. The application of this method for massive particles was possible due to the use of Jacobi's metric. We computed the deflection of massive particles and light for five wormhole backgrounds. The first Casimir wormhole analyzed was built by Garattini \cite{Garattini:2019ivd}, in which the source for the traversable wormhole is the Casimir energy. The other wormholes consider the high energy effects of GUP in this Casimir wormhole. These traversable wormholes with GUP corrections differ by the redshift functions in the metric. We analyzed the deflection of massive particles and light for the redshift functions $\Phi(r) = const$, $\Phi(r) = r_0/r $, $\exp{(2\Phi(r))}=1 + \gamma^2/r^2 $ and $\exp{(2\Phi(r))} = (1 + \beta D_i/r^2)^{-(2/(1+1/\omega))}$.  For each wormhole background, we also investigated the deflection angle of massive particles and light when both source and receiver are very far from the lens ($bu_S \approx bu_R \ll 1$), and always that it is possible, we compared with recent results found in Ref. \cite{Jusufi:2020rpw}

We decided to present our results as a function of the asymptotic velocity $v_\infty$ due to two motivations. The first and more important is that when we represent the gravitational deflection angle of massive particles by its asymptotic velocity we can find the deflection angle of light by doing $v_\infty = 1$. Moreover, if we want to find the deflection as a function of the velocity seen by the receiver, we can use equation (\ref{velocity}).The second motivation is due to all articles which we saw adopt this approach.

First, we found the deflection angle of particles with finite distances for the Garattini background for a constant $\omega$ in Eq. (\ref{EQ23}), give by
\begin{eqnarray}\label{EQ23Conc}
    \alpha &=& \frac{a^2}{b^2}\left[\frac{4{c_1}^2+8c_2+(3{c_1}^2+4c_3)v_\infty^2}{16v_\infty^2}\right]\left[\pi-\arcsin (bu_R) - \arcsin (bu_S)\right]\nonumber \\
    &+& \frac{a}{b}\frac{ c_1(1+v_\infty^2)}{2v_\infty^2}\left[\sqrt{1-b^2{u_R}^2}+\sqrt{1-b^2{u_S}^2}~\right]\nonumber \\
    &-& \frac{a^2{c_1}^2}{b^2}\left[\frac{bu_R(1+v_\infty^2 - b^2{u_R}^2v_\infty^2)^2-b^3{u_R}^3(2b^2{u_R}^2-3)}{8v_\infty^4(1-b^2{u_R}^2)^{3/2}}+\frac{bu_S(1+v_\infty^2 - b^2{u_S}^2v_\infty^2)^2-b^3{u_R}^3(2b^2{u_R}^2-3)}{8v_\infty^4(1-b^2{u_S}^2)^{3/2}}\right] \nonumber \\
    &+& \frac{a^2}{b^2}\left[\frac{4c_3-3{c_1}^2}{64}\right]\left[\frac{bu_R(3-4b^2{u_R}^2)}{\sqrt{1-b^2{u_R}^2}}+\frac{bu_S(3-4b^2{u_S}^2)}{\sqrt{1-b^2{u_S}^2}}\right] + O(a^3/b^3).
\end{eqnarray}
This expression is very general, and from it, we can consider many special cases. For example, we can find the deflection angle of light when considering $v_\infty = 1$ in Eq. (\ref{EQ23Conc}). From the above expression we computed explicitly all the possible cases. When both source and receiver are very far from the lens, the deflection angle of massive particles becomes Eq. (\ref{EQ24}), and the deflection angle of light becomes Eq. (\ref{EQ25}). Moreover, we analyzed the special case $\omega=3$, in which the deflection angle of massive particles with finite distances becomes Eq. (\ref{EQ231}). In addition to it, when both source and receiver are very far from the lens, the deflection angles of massive particles and light are given by Eq. (\ref{EQ26}) and Eq. (\ref{EQ27}), respectively. Of particular importance is the result found in Eq. (\ref{EQ27}), given by
\begin{eqnarray}\label{EQ27Conc}
    {\alpha_\infty}_{[light]} &\approx& \frac{4a}{3b}+\frac{a^2\pi}{3b^2}.
\end{eqnarray}
The second term of the above expression disagrees of the result obtained previously in the literature \cite{Javed:2020mjb}. We believe that this disagreement occurs due to the fact that Ref. \cite{Javed:2020mjb} considers just the trajectory as $u= \frac1b \sin\phi$.

We also found the deflection angle of massive particles with finite distances for the traversable wormholes corrected by GUP. We investigated the background with $\Phi(r) = const$, namely, of zero tidal wormholes, with the corresponding deflection angle given by 
\begin{eqnarray}\label{EQ40Conc}
\alpha &=& \frac{\pi \epsilon}{4b^2} - [\arcsin(bu_R) +\arcsin(bu_S)]\frac{\epsilon}{4b^2}\nonumber \\
&+& \frac{M}{2b}\left[\sqrt{1-b^2u_R^2}+ \sqrt{1-b^2u_S^2}~\right] \nonumber \\
&+&\frac{\epsilon}{4b^2}\left[\frac{bu_R(3-4b^2u_R^2)}{\sqrt{1-b^2u_R^2}}+\frac{bu_S(3-4b^2u_S^2)}{\sqrt{1-b^2u_S^2}}\right]+ O(M\epsilon, M^2, \epsilon^2).
\end{eqnarray}
The interesting point about the above expression is that it does not depends on $v_\infty$. Therefore the deflection angles for massive and massless particles are the same. This is due to the fact that, when $A(r)= constant$ in equation SSS, the deflection of massive particles equals the deflection of light. In the limit case of both source and receiver are very far from the lens ($bu_S \approx bu_R \ll 1$), the deflection angles of massive particles and light become given by Eq. (\ref{EQ41}). This agrees with recent result of Ref. \cite{Jusufi:2020rpw}

Next, we investigated the background with $\Phi(r) = r_0/r$. The deflection angle for this background was presented in Eq. (\ref{EQ50}) and is given by
\begin{eqnarray}\label{EQ50Conc}
\alpha &=&\frac{\epsilon \pi}{4b^2} -\frac{\epsilon}{4b^2}\left[\arcsin(bu_R)+\arcsin(bu_S)\right]+ \left[\frac{M}{2b}- \frac{r_0}{bv_\infty^2}\right] \left[\sqrt{1-b^2u_R^2}+\sqrt{1-b^2u_S^2}\right] \nonumber \\
&+&\frac{\epsilon}{16b^2}\left[\frac{(3-4b^2u_R^2)bu_R}{\sqrt{1-b^2u_R^2}}+\frac{(3-4b^2u_S^2)bu_S}{\sqrt{1-b^2{u_S}^2}}\right]+ O\left(Mr_0,M\epsilon,r_0\epsilon,M^2,r_0^2\right).
\end{eqnarray}
Again, from the above expression we can  find the deflection angle of light when $v_\infty=1$. In the limit case that both source and receiver are very far from the lens ($bu_S \approx bu_R \ll 1$), the deflection angle of massive particles and light becomes Eq. (\ref{EQ51}) and Eq. (\ref{EQ52}), respectively.

The wormhole backgrounds defined by $\exp{(2\Phi(r))}=1 + \gamma^2/r^2 $ and $\exp{(2\Phi(r))} = (1 + \beta D_i/r^2)^{-(2/(1+1/\omega))}$ have similar procedures to compute the deflection angle of massive particles in first order. First, we found the deflection with finite distances by $\exp{(2\Phi(r))}=1 + \gamma^2/r^2$ background in Eq. (\ref{EQ60}). It is given by
\begin{eqnarray}\label{EQ60Conc}
\alpha &=&\frac{\epsilon \pi}{4b^2} +\frac{\pi\gamma^2}{2b^2v_\infty^2}-\left(\frac{\epsilon}{4b^2}+\frac{\gamma^2}{2b^2v_\infty^2}\right)\left[\arcsin(bu_R)+\arcsin(bu_S)\right] \nonumber \\
&+& \frac{M}{2b}\left[\sqrt{1-b^2u_R^2}+\sqrt{1-b^2u_S^2}\right] +\frac{\gamma^2}{2b^2v_\infty^2}\left[\frac{b^3u_R^3}{\sqrt{1-b^2u_R^2}}+\frac{b^3u_S^3}{\sqrt{1-b^2u_S^2}}\right] \nonumber \\
&+&\frac{\epsilon}{16b^2}\left[\frac{(3-4b^2u_R^2)bu_R}{\sqrt{1-b^2u_R^2}}+\frac{(3-4b^2u_S^2)bu_S}{\sqrt{1-b^2{u_S}^2}}\right]+  O\left(M\gamma^2,M\epsilon,\gamma^2\epsilon,M^2,\gamma^4\right).
\end{eqnarray}
The deflection angle of light can be computed by using $v_\infty=1$ in the above expression. Moreover, we found the deflection of massive particles for this background in the case that both source and receiver are very far from the lens in Eqs. (\ref{EQ61}) and the deflection of light can be computed by using $v_\infty=1$ in Eq. (\ref{EQ61}).

Finally, we investigated the deflection of massive particles for $\exp{(2\Phi(r))} = (1 + \beta D_i/r^2)^{-(2/(1+1/\omega))}$ background. The deflection angle of massive particles with finite distances was presented in Eq. (\ref{EQ-5-4-5}), and given by
\begin{eqnarray}\label{EQ-5-4-5Conc}
\alpha &=&\frac{\epsilon \pi}{4b^2} - \frac{1}{1+\omega^{-1}} \frac{\pi\beta D_i}{b^2v_\infty^2}-\left(\frac{\epsilon}{4b^2}-\frac{1}{1+\omega^{-1}} \frac{\beta D_i}{b^2v_\infty^2}\right)\left[\arcsin(bu_R)+\arcsin(bu_S)\right] \nonumber \\
&+& \frac{M}{2b}\left[\sqrt{1-b^2u_R^2}+\sqrt{1-b^2u_S^2}\right] -\frac{1}{1+\omega^{-1}} \frac{\beta D_i}{b^2v_\infty^2}\left[\frac{b^3u_R^3}{\sqrt{1-b^2u_R^2}}+\frac{b^3u_S^3}{\sqrt{1-b^2u_S^2}}\right] \nonumber \\
&+&\frac{\epsilon}{16b^2}\left[\frac{(3-4b^2u_R^2)bu_R}{\sqrt{1-b^2u_R^2}}+\frac{(3-4b^2u_S^2)bu_S}{\sqrt{1-b^2{u_S}^2}}\right]+  O\left(M\gamma^2,M\epsilon,\gamma^2\epsilon,M^2,\gamma^4\right).
\end{eqnarray}
The deflection angle of light can be obtained from the above expression by considering $v_\infty=1$. We also examined the limit when both the source and the receiver are very far from the lens. In this limit case, the deflection angle of massive particles and light become Eq. (\ref{EQ-5-4-6}) and (\ref{EQ-5-4-7}).

\begin{table}[h!]
    \caption{Optical scalars of Casimir wormholes}
    \center
    \label{table1}
\begin{tabular}{lcl}
    & Convergence, $\kappa$ & Shear, $\tilde{\gamma}$ \\
    \hline\hline \\
    GW${^a}$& $- \frac{\lambda_{ls}}{\lambda_s\lambda^2_{l}} \frac{\pi a^2}{6}\frac{1}{\vartheta^3}$ & $\frac{\lambda_{ls}}{\lambda_s}\left(\frac{4a}{3\lambda_l}\frac{1}{\vartheta^2}+\frac{\pi a^2}{2\lambda_l^2}\frac{1}{\vartheta^3}\right)$ \\ \\ \hline \\
    $\Phi = const$& $-\frac{\lambda_{ls}}{\lambda_s\lambda^2_{l}} \frac{\pi^4}{720}\frac{1}{\vartheta^3}$ & $\frac{\lambda_{ls}}{\lambda_s}\left(\frac{1}{\lambda_l}\left(r_0 - \frac{\pi}{90 r_0}\right)\frac{1}{\vartheta^2}+\frac{1}{\lambda_l^2}\frac{\pi^4}{240}\frac{1}{\vartheta^3}\right)$ \\ \\\hline \\
    $\Phi = r_0/r$& $\frac{\lambda_{ls}}{\lambda_s\lambda^2_{l}}\frac{\pi^4}{720} \frac{1}{\vartheta^3}$ & $\frac{\lambda_{ls}}{\lambda_s}\left(\frac{1}{\lambda_l}\left(r_0 + \frac{\pi}{90 r_0}\right)\frac{1}{\vartheta^2}-\frac{1}{\lambda_l^2}\frac{\pi^4}{240}\frac{1}{\vartheta^3}\right)$ \\ \\\hline \\
    $e^{2\Phi} =1 + \gamma^2/r^2 $& $- \frac{\lambda_{ls}}{2\lambda_s\lambda^2_{l}} \left(\frac{\pi^4}{360}+\frac{\pi \gamma^2}{2}\right)\frac{1}{\vartheta^3}$ & $\frac{\lambda_{ls}}{\lambda_s}\left(\frac{1}{\lambda_l}\left(r_0 - \frac{\pi}{90 r_0}\right)\frac{1}{\vartheta^2}+\frac{3}{2\lambda_l^2}\left(\frac{\pi^4}{360}+\frac{\pi \gamma^2}{2}\right)\frac{1}{\vartheta^3}\right)$  \\ \\ \hline \\
    $\omega_r(r) = const$& $- \frac{\lambda_{ls}}{2\lambda_s\lambda^2_{l}} \left(\frac{\pi^4}{360}-\frac{\pi\omega}{1+\omega}\beta D_i\right)\frac{1}{\vartheta^3}$ & $ \frac{\lambda_{ls}}{\lambda_s}\left(\frac{1}{\lambda_l}\left(r_0 - \frac{\pi}{90 r_0}\right)\frac{1}{\vartheta^2}+\frac{3}{2\lambda_l^2}\left(\frac{\pi^4}{360}-\frac{\pi\omega}{1+\omega}\beta D_i\right)\frac{1}{\vartheta^3}\right)$ \\ \\
\end{tabular}\\
\vspace{0.2cm}
${}^a$ \small{GW is the Garattini wormhole, the line element is given by equation (\ref{EQ02})\\and the optical scalar of rotation is null, $\hat{\omega}=0$.}
\end{table}

In this work, we also computed, for each Casimir wormhole, the optical scalars: convergence, shear and rotation represented by $\kappa$, $\tilde{\gamma}$ and $\hat{\omega}$, respectively. We summarize these results in table \ref{table1}.
\section*{Acknowledgement}
The authors would like to thanks Alexandra Elbakyan and sci-hub, for removing all barriers
in the way of science. We acknowledge the financial support provided by the Conselho Nacional de Desenvolvimento Científico e Tecnológico (CNPq), the Coordenação de Aperfeiçoamento de Pessoal de Nível Superior (CAPES) and Fundaçao Cearense de Apoio ao Desenvolvimento Científico e
Tecnológico (FUNCAP) through PRONEM PNE0112- 00085.01.00/16.

\end{document}